\documentclass{article}

\usepackage[numbers]{natbib}         
\usepackage[colorlinks]{hyperref}    
\usepackage[english]{babel} 
\usepackage{amssymb}
\usepackage{amsmath}
\usepackage{txfonts}
\usepackage{mathdots}
\usepackage[classicReIm]{kpfonts}
\usepackage[dvips]{graphicx} 
\usepackage[a4paper, portrait, margin=1in]{geometry}
\usepackage{tabularx}
\usepackage{longtable}
\usepackage{multirow}
\begin{document}


\textbf{\Large Deep Learning in Multi-organ Segmentation}

\textbf{ }

Yang Lei${}^{1}$, Yabo Fu${}^{1}$, Tonghe Wang${}^{1,2}$, Richard L.J. Qiu${}^{1,2}$, Walter J. Curran${}^{1,2}$,

Tian Liu${}^{1,2}$ and Xiaofeng Yang${}^{1,2}$

\bigbreak
${}^{1}$Department of Radiation Oncology, Emory University, Atlanta, GA 

${}^{2}$Winship Cancer Institute, Emory University, Atlanta, GA

\noindent 
\bigbreak
\bigbreak
\bigbreak

\textbf{Corresponding author: }

Xiaofeng Yang, PhD

Department of Radiation Oncology

Emory University School of Medicine

1365 Clifton Road NE

Atlanta, GA 30322

E-mail: xiaofeng.yang@emory.edu

\bigbreak
\bigbreak
\bigbreak
\bigbreak
\bigbreak
\bigbreak

\textbf{Abstract}

This paper presents a review of deep learning (DL) in multi-organ segmentation. We summarized the latest DL-based methods for medical image segmentation and their applications. These methods were classified into six categories according to their network design. For each category, we listed the surveyed works, highlighted important contributions and identified specific challenges. Following the detailed review of each category, we briefly discussed its achievements, shortcomings and future potentials. We provided a comprehensive comparison among DL-based methods for thoracic and head \& neck multi-organ segmentation using benchmark datasets, including the 2017 AAPM Thoracic Auto-segmentation Challenge datasets and 2015 MICCAI Head Neck Auto-Segmentation Challenge datasets.

\noindent \eject 

\noindent 
\section{ INTRODUCTION}

Machine learning (ML) is a hot topic in image processing and has recently gained a lot of interest in medicine \cite{RN1, RN4, RN_Yang}. Artificial Neural Network (ANN) is a subfield of ML that mimic the organization of the brain by using layers of neurons with different weights and biases \cite{RN2, RN6, RN3, RN10, RN8, RN5}. Deep Learning (DL) is a new term for ANN arising from advances in the ANN architectures and algorithms since 2006, referring especially to ANN with many hidden layers. Since there is no consensus on how many layers count as deep, the distinction between ANN and DL is not clearly defined \cite{RN11}. DL has demonstrated enormous potential in computer vision \cite{RN64}. This data-driven method explores vast image features to facilitate various vision tasks, such as image classification \cite{RN65}, object detection \cite{RN67} and segmentation \cite{RN81}. Inspired by the success of DL in computer vision, researchers have attempted to extend the DL-based techniques to medical imaging. DL-based methods have been extensively explored in medical imaging for the purposes of segmentation \cite{RN4, RN31, RN12, RN7, RN8, RN28, RN3, RN43, RN36, RN15, RN34, RN42, RN37, RN51, RN67, RN44, RN39, RN63, RN13}, synthesis \cite{RN9, RN23, RN11, RN38, RN22, RN10, RN19, RN47, RN14, RN18, RN16, RN54, RN17}, enhancement and correction \cite{RN2, RN21, RN45, RN57, RN33, RN13, RN53, RN68, RN69}, and registration \cite{RN1, RN14, RN12, RN16, RN15}. DL-based multi-organ segmentation techniques represent a significant innovation in daily practices of radiation therapy, expediting the segmentation process, enhancing contour consistency and promoting compliance to delineation guidelines \cite{RN4, RN31, RN23, RN22, RN20, RN19, RN18}. Furthermore, rapid DL-based multi-organ segmentation could facilitate online adaptive radiotherapy to improve clinical outcomes.

Lung cancer is the second most common cancer, and the leading cause of cancer death for both male and female \cite{RN33, RN31, RN25, RN26, RN24, RN34}. Depending on the stage and cancer type, 30-60\% of lung cancer patients receive radiation therapy during their treatment \cite{RN24}. Radiotherapy is also the standard care for certain lung cancers \cite{RN35}. The success of radiotherapy depends highly on the control of radiation exposure to the target and organs at risk (OARs), such as lung, esophagus, spinal cord and heart, etc \cite{RN37, RN44}. Therefore, accurate tissue delineation is crucial for the outcome of radiotherapy, especially for highly conformal radiotherapy such as intensity modulated radiotherapy (IMRT), proton therapy and stereotactic body radiotherapy (SBRT). These highly conformal treatments are designed to shape radiation to target volume while sparing healthy OARs, and are usually planned with sharp dose drop-off. Misdelineation could result in severe OARs overdosing. In current clinical practice, targets and OARs are normally delineated manually by clinicians on CT images, which is tedious, time consuming and laborious. CT images provide accurate anatomical information and electron density for treatment planning but are of low soft tissue contrast. This makes the manual delineation of soft tissues, such as the esophagus, particularly difficult and prone to errors arising from inter- and intra-observer variability \cite{RN62, RN58, RN63, RN59, RN60, RN61}. In the last few decades, researchers and clinicians have spent enormous effort to develop automatic contouring methods to provide accurate and consistent organ delineation.

Atlas-based method is a straightforward approach for automatic segmentation, which is available in several commercial products \cite{RN83, RN84, RN82}. This method registers atlas templates that contain pre-contoured structures, to the images to be segmented, and the pre-contoured structures are then propagated to the new images. The segmentation accuracy of this technique depends highly on the accuracy of image registration. Because of organ morphological variabilities across patients and image artifacts, accurate registration is not always guaranteed. This issue can be alleviated by using multi-atlas based methods with a large number  and a wide variety of atlas datasets. However, the unpredictability of tumor shape makes it difficult to include all possible cases in the templates. Moreover, deformable image registration is computational demanding, and a large pool of atlas templates can result in skyrocketed computational cost. Model-based methods make use of statistical shape models for automated segmentation \cite{RN85, RN86, RN87}. The accuracy of those methods depends on the reliability of the models. While models are built based on anatomical knowledge of established datasets, the generalized models have only shown limited success on irregular images.

Thanks to its significant advancements in computer vision, DL-based methods have achieved the-state-of-art performances in medical image segmentation, especially in multi-organ segmentation. In contrast to traditional methods that ultilize handcrafted features, DL-based methods adaptively explore deep features from medical images to represent the image structural information\cite{RN_Wang}. In this review, we summarized popular deep learning-based methods for multi-organ segmentation and discussed the performances of different network designs and the challenges for different segmentation tasks. In addition, since it is easy to modify recent single organ DL-based organ segmention to the multi-organ segmentation method, e.g., change two-channel output to a multi-channel output, we also included several recent single organ segmentation methods. With this survey, we aim to 
\begin{enumerate}
\item  Summarize the latest developments in DL-based medical image multi-organ segmentation.

\item  Highlight contributions, identify challenges and outline future trends.

\item  Provide benchmark evaluations of recently published DL-based multi-organ segmentation methods.
\end{enumerate}

\noindent 
\bigbreak
\section{Data Pre/Post-Processing}

Data directly collected from clinical databases are usually not ready for network training. It is important to perform data preprocessing such as network input size selection, image cropping, image normalization and data augmentation prior to network training. Post-processing techniques are also widely used to refine the network segmentation accuracy. 

\noindent 
\bigbreak
\subsection{Network input sizes}

For multi-organ segmentation, 3D medical image is usually taken into consideration. Depending on the network design and GPU memory limitation, some methods directly use the whole volume as input to train the network \cite{RN91}, while some methods process the 3D image slice by slice \cite{RN90}. Alternatively, 2D/ 3D image patches were used as input in other methods \cite{RN97, RN103}. The 3D-based approaches take 3D patches or whole volume as input and utilize 3D convolution kernels to extract spatial and contextual information from the input images. Full-sized whole volume training often leads to increasing computational cost and complexity as increasing number of layers are used. Compared to whole volume-based methods, some methods extract 3D small patches from 3D image by sliding window across original images prior to network prediction, and then use patch fusion to obtain the final full-sized segmentation. 3D patch-based methods are less computational and GPU-memory demanding. For example, Vaidhya \textit{et al.} extracted 3D patches from MRI to segment brain Gliomas tumor \cite{RN103}. 
The 2D-based approaches take the full-sized transverse slices or 2D patches from transverse, sagittal or coronal view as input and feed them into the 2D network consists of 2D convolutional kernels. The 2D-based approaches can easily and efficiently reduce the memory requirement and computational cost, however, it ignores the spatial information of the medical image in the third dimension. For example, in U-Net method, Ronneberger \textit{et al.} extracted 2D slices from electron microscopic image and fed them into U-Net model to segment the neural structure \cite{RN90}. In addition, to exploit the 3D feature information while using 2D kernels, Lei \textit{et al.} proposed a multidirectional method to extract features from axial, coronal and sagittal views separately and then used a surface-based contour refinement to combine the segmented contours from the three dimensions \cite{RN28}. The method can learn rich feature information in three dimensions and achieve good performance for prostate segmentation.

\noindent 
\bigbreak
\subsection{Pre-processing}

Pre-processing plays an important role in segmentation tasks, since there are variant intensity, contrast and noise in the images. To ease the network training, pre-processing techniques are usually applied prior to network training. Typical pre-processing techniques include registration \cite{RN1}, bias/scatter/attenuation correction \cite{RN49, RN50}, voxel intensity normalization \cite{RN243} and cropping \cite{RN44} etc.

\noindent 
\bigbreak
\subsection{Data augmentation}

Data augmentation is used to reduce over-fitting and increase the amount of training samples. Typical data augmentation techniques include rotation, translation, scaling, flipping, distortion, linear warping, elastic deformation, and noise contamination.

\noindent

\noindent 
\bigbreak
\subsection{Post-processing}

Post-processing is applied to refine the segmented contours to be more smooth, continuous and realistic. Generally, morphological techniques are used to remove unreal binary masks of contours. Recently some post-processing methods are developed according to the structure of segmented contours. Conditional random field is one of the most popular post-processing methods in deep learning-based segmentation methods \cite{RN91}.

\noindent

\noindent 
\bigbreak
\section{Deep Learning in Medical Image Multi-Organ Segmentation}

The task of medical image multi-organ segmentation is typically defined as assigning each voxel of the medical images to one of several labels that represent the objects of interest. Segmentation is one of the most commonly studied DL-based applications in the medical field. Therefore, there are a wide variety of methodologies with many different network architectures.
DL-based multi-organ segmentation methods can be divided from different aspects according to its properties such as network architecture, training process (supervised, semi-supervised, unsupervised, transfer learning), input size (patch-based, whole volume-based, 2D, 3D) and so on. In this paper, we classify them according to its architecture into 6 categoreis, including 1) auto-encoder (AE), 2) convolutional neural network (CNN), 3) fully convolutional network (FCN), 4) generative adversarial network (GAN), 5) regional convolutional neural network (R-CNN) and 6) hybrid DL-based methods. In each category, we provided a comprehensive table, listing all the surveyed works belonging to this category and summarizing their important features. Besides multi-organ segmentation methods, we also included single-organ segmentation methods and object detection methods, and summarized these methods from the network design perspective.  Single-organ segmentation can be easily transformed tomulti-organ segmentitation by replacing the last layer’s binary output to a multi-channel binary output. Medical image object detection methods were included since they could be used to obtain the region of interest (ROI) to aid the segmentation procedure and improve the segmentation accuracy.
Before we delve into the details of each category, we provide a detailed overview of DL-based medical image multi-organ segmentation methods with their corresponding components and features in Fig. 1. The aim of Fig. 1 is to give the readers an overall understanding of each category by putting its important features side by side with each other.

\begin{figure}
\centering
\noindent \includegraphics*[width=6.50in, height=4.20in, keepaspectratio=true]{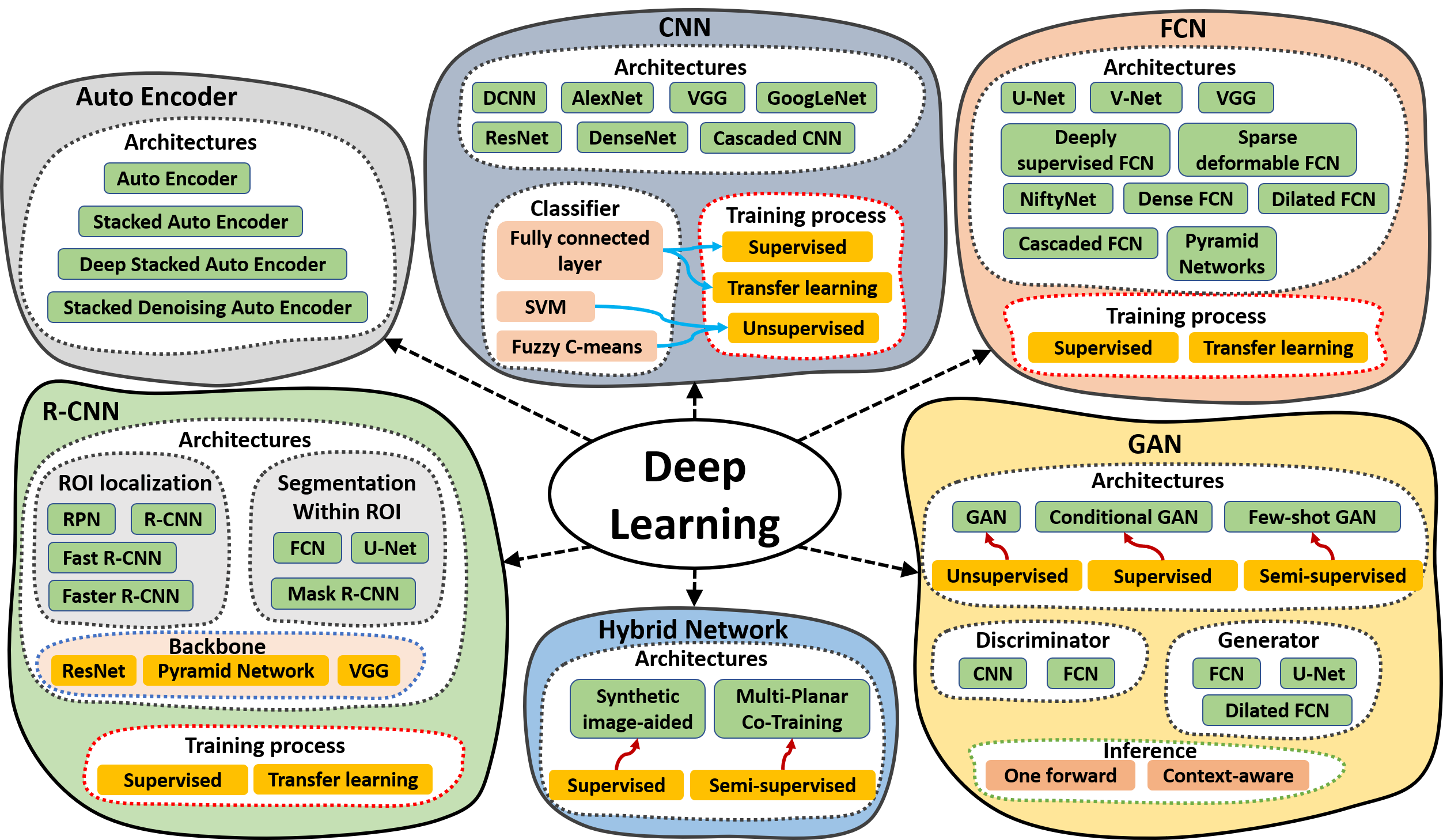}

\noindent Fig. 1. Overview of six categories of DL-based methods in medical image segmentation.
\end{figure}

Works cited in this review were collected from various databases, including Google Scholar, PubMed, Web of Science, Semantic Scholar and so on. To collect as many works as possible, we used a variety of keywords including but not limited to deep learning, multi-organ, medical segmentation, convolutional neural network and so on. We totally collected over 180 papers that are closely related to DL-based medical image segmentation and over 40 papers that are closely related to multi-organ segmentation. Most of these works were published between the year of 2017 and 2019. The number of multi-organs publications is plotted against year by stacked bar charts in Fig. 2. Number of papers were counted by six categories. The dotted line of Fig. 2 indicates increase interest in DL-based multi-organ segmentation methods over the years. Especially, the total number of publications has grown dramatically over the last 2 years. 

\begin{figure}
\centering
\noindent \includegraphics*[width=4.77in, height=3.00in, keepaspectratio=true]{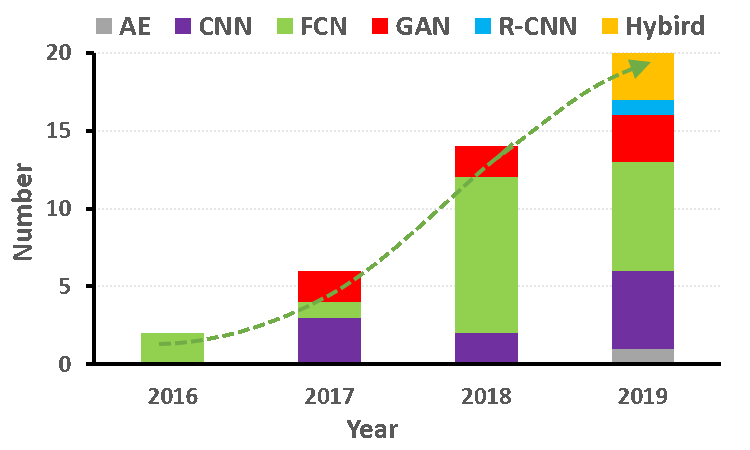}

\noindent Fig. 2. Overview of number of publications in DL-based multi-organ segmentation.
\end{figure}

\noindent 
\bigbreak
\subsection{Auto-encoder methods}
\noindent 
\bigbreak
{\bf 3.1.1 Auto-encoder and its variants}

In the literature, AE and its variants have been extensively studied and continue to be utilized in medical image analysis \cite{RN99}. AE usually consists a single neural network encoder layer that transforms the input into a latent or compressed representation by minimizing the reconstruction errors between input and output values of the network, and a single neural network decoder layer that restores the original input from the low-dimensional latent space. By constraining the dimension of latent representation, AE can discover relevant pattern from the data. 
To prevent an AE from learning an identity function, several improved AEs were proposed. The most popular and widely used network model in deep unsupervised architecture is stacked AE (SAE). SAE is constructed by organizing AEs on top of each other also known as deep AEs. SAEs consist of multiple AEs stacked into multiple layers where the output of each layer is wired to the inputs of the successive layers \cite{RN96}. To obtain good parameters, SAE uses greedy layer-wise training. The benefit of SAE is that it can benefit from deeper network, which has greater expressive power. Furthermore, it usually captures a useful hierarchical grouping of the input \cite{RN96}. 
Denoising autoencoder (DAEs) is another variant of the AE and used to constitute better higher-level representation and extract useful features \cite{RN101}. DAEs prevent the model from learning a trivial solution where the model is trained to reconstruct a clean input from the corrupted version from noise or another corruption \cite{RN97}. Stack denoising autoencoder (SDAE) is a deep network utilizing the power of DAE \cite{RN103}.
The limitation of AEs to have only small numbers of hidden units can be overcome by adding a sparsity constraint. The aim of sparse autoencoder is to make a large number of neurons to have low average output so that neurons may be inactive most of the time \cite{RN98}. Sparsity can be achieved by introducing a loss function during training or manually zeroing a few strongest hidden unit activations.
Stacked AE requires layer-wise pre-training. When layers go deeper during the pre-training process, it may be time consuming and tedious because of stacked AE is built with fully connected layers. Li \textit{et al.} proposed first trial to train convolutional directly an end-to-end manner without pre-training \cite{RN104}. Guo \textit{et al.} suggested convolutional autoencoder (CAE) that is beneficial to learn feature for images, preserve local structures and avoid distortion of feature space \cite{RN105}. Wang \textit{et al.} proposed an automated chest screening based on a hybrid model of transfer learning and CAE \cite{RN100}.

\noindent 
\bigbreak
{\bf 3.1.2 Overview of works}

Since abnormalities, e.g., abnormal tissue types and irregular organ shapes are often present in medical images, it is challenging to obtain ground truth labels of multi-organs for supervised learning. However, organ segmentation in such abnormal dataset is meaningful in automated radiotherapy. Shin \textit{et al.} applied a SAE method for organ detection in magnetic resonance imaging (MRI) \cite{RN96}. Their method was used to detect the locations of the liver, heart, kidney and spleen for MRI scans of the abdominal region containing liver or kidney metastatic tumors. Only weakly supervised training is required to learn visual and temporal hierarchical features that represent object classes from an unlabeled multimodal dynamic contrast-enhanced magnetic resonance imaging (DCE-MRI) data. A probabilistic patch-based method was employed for multiple organ detection, with the features learned from the SAE model.
Accurate and automated segmentation of glioma from MRI is important for treatment planning and monitoring disease progression. Vaidhya \textit{et al.} used SDAE to solve the challenge of variable shape and texture of glioma tissue in MRI for this segmentation task \cite{RN103}. 3D patches were extracted from multiple MRI sequences then were fed into the SDAE model to obtain the glioma segmentation. During training, two SDAE models were supervised in this task, one for high grade glioma (HGG) data, the other one for a combination of HGG and low-grade glioma (LGG) data. During testing, the segmentation was obtained by a combination of predictions from two trained networks via maximum a posteriori (MAP) estimation. Simultaneously, Alex \textit{et al.} applied SDAE for brain lesion detection, segmentation, and false-positive reduction \cite{RN97}. SDAE was pretrained using a large number of unlabeled patient volumes and fine-tuned with 2D patches drawn from a limited number of patients. LGG segmentation was achieved using a transfer learning approach in which a SDAE network pretrained with LGG data was fine-tuned using LGG data. 
Ahmad \textit{et al.} proposed a deep SAE (DSAE) for CT liver segmentation \cite{RN107}. First, deep features were extracted from unlabeled data using the AE. Second, these features are fine-tuned to classify the liver among other abdominal organs.
In order to efficiently detect and identify normal levels during mass chest screening of lung lesions of chest X-rays (CXRs), Wang \textit{et al.} proposed a convolutional SDAE (CSDAE) to determine which three levels of the images (i.e. normal, abnormal, and uncertain cases) the CXRs fall into \cite{RN100}.
Accurate vertebrae segmentation in the spine is essential for spine assessment, surgical planning and clinical diagnostic treatment. Qadri \textit{et al.} proposed a stacked SAE (SSAE) model for the segmentation of vertebrae from CT images \cite{RN98}. High-level features were extracted via feeding 2D patches into the SSAE model in an unsupervised way. To improve the discriminability of these features, a further refinement using a supervised fashion and fine-tuning was integrated. Similarly, Wang \textit{et al.} proposed to localize and identify vertebrae by combining SSAE contextual features and structured regression forest (SRF) \cite{RN106}. Contextual features were extracted via SSAE in an unsupervised way, and were then fed into SRF to achieve whole spine localization.

\bigbreak
\noindent \textbf{Table 1} Overview of AE methods

\begin{longtable}{|p{0.3in}|p{0.3in}|p{1.2in}|p{1.0in}|p{0.8in}|p{0.8in}|p{0.6in}|} \hline 
\textbf{Ref.} & \textbf{Year} & \textbf{Network} & \textbf{Supervision} & \textbf{Dimension} & \textbf{Site} & \textbf{Modality} \\ \hline 
\cite{RN96} & 2013 & SAE & Weakly supervised & 3D patch & Abdomen & 4D DCE-MRI \\ \hline 
\cite{RN103} & 2015 & SDAE & Supervised & 3D patch & Brain Gliomas & MRI \\ \hline 
\cite{RN97} & 2017 & SDAE & Semi-supervised & 2D patch & Brain lesion & MRI \\ \hline 
\cite{RN107} & 2017 & SAE & Transfer learning & 2D slice & Liver & CT \\ \hline 
\cite{RN100} & 2018 & CSDAE & Transfer learning & 2D slice & Thoracic & chest X-rays \\ \hline 
\cite{RN98} & 2019 & SSAE & Unsupervised & 2D patch & Vertebrae & CT \\ \hline 
\cite{RN106} & 2019 & SSAE & Unsupervised & 2D patch & Vertebrae & CT \\ \hline 
\cite{RN144} & 2019 & Hierarchical 3D neural networks & Supervised & N.A. & Head \& Neck & CT \\ \hline
\end{longtable}

\noindent 
\bigbreak
{\bf 3.1.3 Discussion}

For the segmentation of public BraTS 2013 and BraTS 2015 data \cite{RN109}, which are multi-modality brain MRI tumor segmentation datasets, SDAE can provide good segmentation performance \cite{RN97}. For segmenting liver on CT images, DSAE showed high classification accuracy and can speed up the clinical task \cite{RN107}.
For detecting and identifying normal levels during mass chest screening of lung lesions of CXRs, CSDAE method achieves promising results in terms of precision of 98.7 and 94.3\% based on the normal and abnormal cases, respectively \cite{RN100}. The results achieved by the proposed framework show superiority in classifying the disease level with high accuracy. CSDAE can potentially save the radiologists time and effort, allowing them to focus on higher-level risk CXRs.
In contrast to machine learning, whose performance depends on hand-craft features, AE can learn medical image deep contextual features instead of hand-crafted ones by building larger-range input samples to improve their contextual discrimination ability \cite{RN106}.
Validated on public MICCAI CS2014 dataset which includes a challenging dataset of 98 spine CT scans, the SSAE method could effectively and automatically locate and identify spinal targets in CT scans, and achieve higher localization accuracy, low model complexity without making any assumptions about visual field in CT scans \cite{RN98}.
Although AE has many benefits, it faces some challenges and limitations in medical multi-organ segmentations. One of the limitations is related to data regularity. For example, in case of anatomical structures like lung, heart, liver, even if the inter-subject variability of dataset is high, the shape variety of segmentation masks would remain low. Unlike organs which tend to have similar structure, irregular lesions and tumors with large shape variability are difficult for AE to encode and remain challenging for the unsupervised AE methods. On the other hand, due to the neural networks used in AE methods, the number of layers would be limited due to large computation complexity. Compared to CNN which uses convolution kernels with shared learnable parameters, AE methods are limited to use large number of layers to learn deeper features.

\noindent 
\bigbreak
\subsection{CNN methods}

\noindent 
\bigbreak
{\bf 3.2.1 Network designs}

Neural network was previously used for medical image segmentation \cite{RN169}. CNN is one of the variants of neural network used in the field of computer vision and medical imaging. It derives its name from the type of hidden layers it consists of. The hidden layers of a CNN typically consist of convolutional layers, max pooling layers, batch normalization layers, dropout layers and fully connected layers, and normalization layers. The last layer of a CNN is typically a sigmoid or softmax layer for classification and tanh layer for regression.
Convolution layers are the core of CNN and are used for feature extraction \cite{RN13}. The convolution layer extracts variant feature maps depending on its learned convolution kernels. Pooling layer performs a down-sampling operation by using maximum or average of the defined neighborhood as the value to reduce the spatial size of each feature map. Rectified linear unit (ReLU) and its modifications such as Leaky ReLU are among the most commonly used activation functions, which transforms data by clipping any negative input values to zero while positive input values are passed as output \cite{RN207}. Fully connected layer connects every neuron in previous layer to every neuron in next layer. Neurons in a fully connected layer are fully connected to all activations in the previous layer. They are placed before the classification output of a CNN and are used to flatten the results before a prediction is made using linear classifiers. Via several fully connected layers, the previous feature maps extracted from convolutional layers are represented to probabilities to classify the medical image or medical image patch or voxels. 
During training the learnable parameters of CNN architecture, the model predicts the class scores for training images, computes the loss using the selected loss function and finally updates the weights using the gradient descent method by back-propagation. Cross-entropy loss is one of the most widely used loss functions and stochastic gradient descent (SGD) and Adam gradient descent optimizations are the most popular method to operate gradient descent.
Lecun \textit{et al.} first proposed a CNN model, named by LeNet, for hand-written digit recognition \cite{RN209}. LeNet is composed of convolution layers, pooling layers and full connected layers. With the development of computer hardware and the increase in the amount of data available for neural network training, in 2012, Krizhevsky \textit{et al.} proposed AlexNet and won the ILSVRC-2012 image classification competition \cite{RN210} with a far lower error rate than the second place \cite{RN65}. Since then, CNNs have begun to attract widespread attention, and variants of CNNs have been developed and have achieved the-state-of-art performances in various image processing tasks. The improvements of AlexNet as compared to LeNet include 1) ReLU layer for nonlinearity and sparsity, 2) data augmentation to enlarge the dataset variety, 3) dropout layer to reduce learnable parameters and prevent overfitting, 4) GPU for parallel computing, 5) local response normalization to normalize the nearest data and 6) overlapping pooling. Additionally, Zeiler and Fergus proposed ZFNet to improve the performance of AlexNet \cite{RN211} and proved that shallow network is able to learn edge, color and texture features of images and high-level network can learn abstract features of images. In addition, they demonstrated better performance can be achieved via deeper network. The main improvement of ZFNet is deconvolution network used to visualize the feature map. 
Simonyan and Zisserman proposed the VGG to further explore the performance of the deeper network model \cite{RN213}. The main innovation of VGG is a thorough evaluation of networks of increasing depth using an architecture with very small (3×3) convolution filters, which shows that a significant improvement on the prior-art configurations can be achieved by pushing the depth to 16 to 19 layers. Similarly, GoogLeNet was proposed to broaden the network structure \cite{RN214}. By integrating proposed inception module, GoogLeNet won the winner of the ImageNet Large-Scale Visual Recognition Challenge 2014 (ILSVRC14), which is an image classification and detection competition. The inception module is helpful for the CNN model to better describe the input data content while further increasing the depth and width of the network model.
The above development of CNNs is to increase the depth and width of CNN to improve the performance. However, simply increasing the depth would lead to vanishing/exploding gradients. To ease the difficulty of training deep CNNs and solve the degradation effect caused by increasing network depth, He \textit{et al.} proposed a residual network (ResNet) for image recognition \cite{RN216}. ResNet, which is mainly composed of residual blocks, is demonstrated to be able to break through a 100-layers barrier and even reach 1000 layers.
Inspired by residual network, Huang \textit{et al.} later proposed a densely connected convolutional network (DenseNet) by connecting each layer to every other layer \cite{RN222}. Contrast to residual block, which would focus on learn the structural difference between the feeding image and learning target, DenseNet aimed to combine both low-frequency and high-frequency feature maps from previous and current convolutional layers via dense blocks.
\noindent

\noindent
\bigbreak 
{\bf 3.2.2 Overview of works}

In medical image segmentation, CNNs can be simply used to classify each voxel or patch in the image individually, by presenting it with patches extracted around that voxel or patch. Roth \textit{et al.} proposed a multi-level deep CNN approach for pancreas segmentation in abdominal CT image \cite{RN89}. A dense local image patch label was obtained by extracting an axial-coronal-sagittal viewed patch in a sliding window manner. The proposed CNN learns to assign class probabilities for each center voxel of its patch. Finally, a stacked CNN leveraged the joint space of CT intensities and dense probability maps. The used CNN architecture consists five convolutional layers which are followed by max-pooling, three fully connected layers, two drop out layer and a soft-max operator to perform binary classification. This CNN architecture can be introduced into multi-organ segmentation frameworks by specifying more tissue types since CNN naturally supports multi-class classifications \cite{RN223}.
In contrast to 2D input, which would lose spatial information, Hamidian \textit{et al.} proposed to use 3D patch-based CNN to detect lung pulmonary nodules for chest CT images \cite{RN167} using volumes of interest extracted from the lung image database consortium (LIDC) dataset \cite{RN224}. They extended previous 2D CNN to three dimensions which would be more suitable for volumetric CT data. 
For high pathologically affected cases, segmenting and classifying the lytic and sclerotic metastatic lesions from CT image is challenging, because these lesions are ill-defined. Therefore, it is hard to extract relevant features that can well-represent texture and shape information for traditional machine learning-based method. In order to solve this problem, Chmelik \textit{et al.} applied deep CNN (DCNN) to segment and classify these kinds of lesions \cite{RN182}. The CNN architecture takes three perpendicular 2D patches for each voxel of 3D CT image as input and output classification of three categories (healthy, lytic and sclerotic) for that voxel. The proposed CNN consisted of several convolutional layers which are followed by ReLU and max-pooling to extract features, several fully connected layers with dropout layers to combine the feature maps to feature vectors, and a last fully connected layer to convert the feature vector to a three-element output of class scores. A high score correlates to a high probability to the corresponding class. L2 regularized cross-entropy and class error loss are used for optimization. Mini-batch gradient descent with momentum back-propagation algorithm is used to optimize the learnable parameters of this CNN.
During radiotherapy for nasopharyngeal carcinoma (NPC) treatment, accurate segmentation of OARs in head and neck (H\&N) CT image is a key step for effective planning. Due to low-contrast and surrounding adhesion tissues of the parotids, thyroids and optic nerves, automatic segmenting these regions is challenging and will result in lower accuracy for these regions as compared to other organs. In order to solve this challenge, Zhong \textit{et al.} proposed a cascaded CNN network to delineate these three OARs for NPC radiotherapy by combining boosting algorithm \cite{RN148}. In their study, CT images of 140 NPC patients treated with radiotherapy were collected. Manual contours of three OARs were used as learning target. A hold-out test was used to evaluate the performance of the proposed method, i.e., the datasets were divided into a training set (100 patients), a validation set (20 patients), and a test set (20 patients). In the boosting method for combining multiple classifiers, three cascaded CNNs for segmentation were combined. The first network was trained with the traditional approach. The second one was trained on patterns (pixels) filtered by the first net. That is, the second machine recognized a mix of patterns (pixels), 50\% of which was accurately identified by the first net. Finally, the third net was trained on the new patterns (pixels) screened jointly by the first and second networks. During the test, the outputs of the three nets were considered to obtain the final output. 2D patch-based ResNet \cite{RN216} was used to build the cascaded CNNs.
For multi-OARs segmentation in thoracic radiotherapy treatment, Harten \textit{et al.} proposed a combination of 2D and 3D CNNs for automatic segmentation of OARs (including esophagus, heart, trachea, and aorta) on thoracic treatment planning CT scans of patients diagnosed with lung, breast or esophageal cancer \cite{RN185}. The two CNNs are summarized as follows: one 3D patch-based network that contains a deep segment of residual blocks \cite{RN225} with sigmoid layer to perform multi-class binary classification, and one 2D patch-based (2D patch extracted from axial, coronal and sagittal planes) network containing dilated convolutions \cite{RN226} with softmax layer to perform classification. A hold-out validation (40 data for training and 20 data for testing) was used to evaluate the performance of the proposed method.

\bigbreak
\noindent \textbf{Table 2} Overview of CNN methods

\begin{longtable}{|p{0.3in}|p{0.3in}|p{1.4in}|p{0.8in}|p{0.8in}|p{0.9in}|p{0.5in}|} \hline 
\textbf{Ref.} & \textbf{Year} & \textbf{Network} & \textbf{Supervision} & \textbf{Dimension} & \textbf{Site} & \textbf{Modality} \\ \hline 
\cite{RN19} & 2017 & Deep deconvolutional neural network (DDNN) & Supervised & 2D slice & Brain & CT \\ \hline 
\cite{RN89} & 2015 & Multi-level DCNN & Supervised & 2D patch & Pancreas & CT \\ \hline 
\cite{RN154} & 2016 & Holistically Nested CNN & Supervised & 2D patch & Pancreas & CT \\ \hline 
\cite{RN167} & 2017 & 3D CNN & Supervised & 3D patch & Chest & CT \\ \hline 
\cite{RN153} & 2017 & 3D DCNN & Supervised & N.A.* & Abdomen & CT \\ \hline 
\cite{RN88} & 2017 & CNN & Supervised & 3D patch & Head \& Neck & CT \\ \hline 
\cite{RN168} & 2017 & Fuzzy-C-Means CNN & Supervised & 3D patch & Lung nodule & CT \\ \hline 
\cite{RN115} & 2017 & DCNN & Supervised & 2D Slice & Body, Chest, Abdomen & CT \\ \hline 
\cite{RN179} & 2018 & Fusion Net & Supervised & 2D patch & 100 ROIs  & HRCT \\ \hline 
\cite{RN182} & 2018 & DCNN & Supervised & 2D patch & Spinal lesion & CT \\ \hline 
\cite{RN178} & 2018 & DCNN & Supervised & 2D slice & Malignant pleural mesothelioma & CT \\ \hline 
\cite{RN166} & 2018 & 2D and 3D CNN & Supervised & 2D slice, 3D volume & Artery / vein & CT \\ \hline 
\cite{RN124} & 2018 & 3D ConvNets & Transfer learning & 3D volume & Brain & MRI \\ \hline 
\cite{RN126} & 2018 & CNN with specific fine-tuning & Supervised or Unsupervised & 2D slice, 3D volume & Brain, abdomen & Fetal MRI \\ \hline 
\cite{RN129} & 2018 & 2D and 3D DCNN & Supervised & 2D slice, 3D volume & Whole body & CT \\ \hline 
\cite{RN160} & 2019 & Deep fusion Network & Supervised & 2D slice & Chest & CXR \\ \hline 
\cite{RN143} & 2019 & DCNN & Supervised & 2D slice & Abdomen & CT \\ \hline 
\cite{RN165} & 2019 & 2.5D CNN & Supervised & 2.5D patch & Thorax & CT \\ \hline 
\cite{RN148} & 2019 & Cascaded CNN & Supervised & 2D slice & Head \& Neck & CT \\ \hline 
\cite{RN185} & 2019 & 2D and 3D CNN & Supervised & 2D slice, 3D volume & Thorax & CT \\ \hline 
\cite{RN180} & 2019 & U-Net Neural Network & Supervised & 3D patch & Lung & CT \\ \hline 
\end{longtable}

*N.A.: not available, i.e. not explicitly indicated in the publication

\noindent 
\bigbreak
{\bf 3.2.3 Discussion}

For the multi-OARs segmentation of CT H\&N, Dice similarity coefficient (DSC), 95th percentile of the Hausdorff distance (95\% HD), and volume overlap error (VOE) were used to assess the performance of cascaded CNN \cite{RN148}. The mean DSC values were above 0.92 for parotids, above 0.92 for thyroids, and above 0.89 for optic nerves. The mean 95\% HDs were approximately 3.08 mm for parotids, 2.64 mm for thyroids, and 2.03 mm for optic nerves. The mean VOE metrics were approximately 14.16\% for parotids, 14.94\% for thyroids, and 19.07\% for optic nerves. From the comparison in \cite{RN148}, the proposed boosting-based cascaded CNN outperformed U-Net \cite{RN90} in segmenting the three OARs. Despite the powerful performance of boosting structure, its pixel-based classification took more time as compared to U-Net. This is because all classifiers in the boosting structure need to classify all pixels in the image.
In the study of \cite{RN185}, researchers evaluated the performance for 2D CNN, 3D CNN and a combination of 2D and 3D CNNs individually and demonstrated the combination network produces the best results. The DSC of the esophagus, heart, trachea, and aorta were 0.84$\mathrm{\pm}$ 0.05, 0.94$\mathrm{\pm}$ 0.02, 0.91$\mathrm{\pm}$ 0.02, and 0.93$\mathrm{\pm}$ 0.01, respectively. These results demonstrate potential for automating segmentation of OARs in routine radiotherapy treatment planning. 
A drawback of most CNN methods is classification need to be performed on every voxel or small patch repeatedly. By sliding window with huge overlap between two neighboring patches, the CNN models perform classifications on each voxel of the whole volume. This approach is inefficient since it requires repeated forward network prediction on every voxel of the image. Fortunately, the convolution and dot product are both linear operators and thus inner products can be written as convolutions and vice versa \cite{RN208}. By rewriting the fully connected layer as convolutions, the traditional CNNs can take input image larger than its training image and produce a likelihood map, rather than an output for a single voxel. However, this may result in output with a far lower resolution than the input due to the pooling layers used. 

\noindent 
\bigbreak
\subsection{FCN methods}
\noindent 
\bigbreak
{\bf 3.3.1 Network designs}

In most CNN methods, the downsized input image or patch goes through the convolutional layers, fully connected layers and output predicted label for the center voxel of the input image. Shelhamer \textit{et al.} first proposed a CNN whose last fully connected layer is replaced by a convolutional layer. Since all layers in this CNN are convolutional layers, the new network is named as FCN. Due to the major improvement of deconvolution kernels used to up-sample the feature map, FCN allows the model to have a dense voxel-wise prediction from the full size whole volume instead of a patch-wise classification as in traditional CNN \cite{RN81}. This segmentation is also called as end-to-end segmentation. By using FCN, the segmentation of whole image can be achieved in just one forward pass. To achieve better localization performance, high-resolution activation maps are combined with up-sampled outputs and then passed to the convolution layers to assemble more accurate output.
One of the most well-known FCN structures for medical image segmentation is U-Net, initially proposed by Ronneberger \textit{et al.} using the concept of deconvolution \cite{RN90}. U-Net architecture is built upon the elegant architecture of FCN, which includes an encoding path and a decoding path. Besides the increased depth of network with 19 layers, U-Net introduced a superior design of long skip connections between the layers of equal resolution in encoding path to decoding path. These connections provide essential high-resolution features to the deconvolution layers. The improvement of U-Net is expected to overcome the trade-off between organ localization and the use of context. This trade-off rises since the large size patches require more pooling layers and consequently will reduce the localization accuracy. On the other hand, small-sized patches can only observe small context of input.
Inspired by the study of U-Net, Milletari \textit{et al.} proposed an improved network based on U-Net, called as V-Net \cite{RN227}. The V-Net architecture is similar to U-Net. It also consists of encoding path (compression) and decoding path and the long skip connection between the encoding and decoding paths. The improvement of V-Net as compared to U-Net is that at each stage of encoding and decoding path, V-Net involves residual block as short skip connection between early and later convolutional layers. This architecture ensures convergence compared with non-residual learning network, such as U-Net. Second, V-Net replaces the max pooling operations with convolutional layers to force the network to have a smaller memory footprint during training, since no switches mapping the output of pooling layers back to their inputs are needed for back propagation. Third, in contrast to binary cross entropy loss used in U-Net method, V-Net used dice loss. Thus, weights to samples of different organs to establish balance between multi-organs and background voxels are not needed.
As another improvement of U-Net, Christ \textit{et al.} proposed a new FCN that by cascading two U-Nets to improve the accuracy of segmentation \cite{RN91}, called as cascade FCN. The main idea of cascade FCN is to stack a series of FCN in the way that each model utilizes the contextual features extracted by the prediction map of the previous model. A simple design is to combine FCNs in a cascade manner, where the first FCN segments the image to ROIs for the second FCN, where the organs segmentation is done. The advantage of using such a design is that separate sets of filters can be applied for each stage and therefore the quality of segmentation can be significantly improved.
The main idea of deep supervision in deeply supervised FCN methods \cite{RN28, RN34} is to provide the direct supervision of the hidden layers and propagate it to lower layers, instead of using only one supervision at the output layer for traditional supervised FCN. In this manner, supervision was extended to deep layers of the network, which would enhance the discriminative ability of feature maps to differentiate multiple classes in multi-organ segmentation task. In addition, recently, attention gate was used in FCN to improve performance in image classification and segmentation \cite{RN228}. Attention gate could learn to suppress irrelevant features and highlight salient features useful for a specific task.

\noindent 
\bigbreak
{\bf 3.3.2 Overview of works}

Multi-organ segmentation aims to segment several organs simultaneously. Zhou \textit{et al.} proposed 2.5D FCN segmentation method to automatically segment 19 organs in CT images of whole body \cite{RN114}. In this work, 2.5D patch, which consists of several consistent slices of axial plane, was used as multi-channel input for the 2D FCN. Separate FCN was designed for each 2D sectional view, resulting in a total of three FCNs. Then, by using fusion method, the final segmentation can be obtained from the three segmentations. The technique produced higher accuracy for big organs such as liver (a Dice value of 0.937) but yielded lower accuracy when dealing with small organs, such as pancreas (a Dice value of 0.553). In addition, by implementing the convolution kernels via a 3D manner, FCN has also been used for multi-organ segmentation for 3D medical images \cite{RN153}.
The FCN which has been trained on the whole 3D images has high class imbalance between the foreground and background, which results into inaccurate segmentation of small organs. One possible solution to alleviate this issue is applying two-step segmentation in a hierarchical manner, where the second stage uses the output of the first stage by focusing more on boundary regions. Christ \textit{et al.} performed liver segmentation by cascading two FCNs, where the first FCN detects the liver location via estimating the ROI, and the second FCN extracts features from that ROI to obtain the liver lesions segmentation \cite{RN91}. This system has achieved 0.823 in Dice for lesion segmentation in CT images and 0.85 in MRI images. Similarly, Wu \textit{et al.} investigated the cascaded FCN to improve the performance in fetal boundary detection in ultrasound images \cite{RN229}. Their results have shown better performance compared to other boundary refinement techniques for ultrasound fetal segmentation.
Transrectal ultrasound (TRUS) is a versatile and real-time imaging modality that is commonly used in image-guided prostate cancer interventions (e.g., biopsy and brachytherapy). Accurate segmentation of the prostate is key to biopsy needle placement, brachytherapy treatment planning, and motion management. However, the TRUS image quality around prostate base and apex region is often affected by low contrast and image noise. To address these challenging, Lei \textit{et al.} proposed a  deep supervision in V-Net for accurate prostate segmentation \cite{RN28}. To cope with the optimization difficulties of training the DL-based model with limited training data, a deep supervision strategy with a hybrid loss function (logistic and Dice loss) was introduced to the different stages of decoding path. To reduce possible segmentation errors at the prostate apex and base in TRUS images, a multi-directional-based contour refinement model was introduced to fuse transverse, sagittal and coronal plane-based segmentation. 
Similarly, for the task of MRI pelvic segmentation, the segmentation of prostate is challenging due to the inhomogeneous intensity distributions and variation in prostate anatomy. Bo \textit{et al.} proposed a 3D FCN with deep supervision and group dilated convolution to segment the prostate on MRI \cite{RN34}. In this method, the deep supervision mechanism was introduced into FCN to effectively alleviate the common exploding or vanishing gradients problems in training deep models, which forces the update process of the hidden layer filters to favor highly discriminative features. A group dilated convolution which aggregates multi-scale contextual information for dense prediction was proposed to enlarge the effective receptive field. In addition, a combined loss (including cosine and cross entropy) was used to improve the segmentation accuracy from the direction of similarity and dissimilarity.
Segmenting glands is essential in cancer diagnosis. However, accurate automated DL-based segmentation of glands is challenging because a large variability in glandular morphology across tissues and pathological subtypes exist, and a large number of accurate gland annotations from several tissue slides is required. Binder \textit{et al.} investigated the idea of  cross-domain (-organ type) approximation that aims at reducing the need for organ-specific annotations \cite{RN130}. Two proposed Dense-U-Nets are trained on Hematoxylin and eosin (H\&E) strained colon adenocarcinoma samples focusing on the gland and stroma segmentation. Unlike U-Net, Dense-U-Nets uses asymmetric encoder and decoder. The encoder is designed to automatically and adaptively learn the spatial hierarchies of features from low to high level patterns coded within the image. The encoder uses transition layer (convolution with stride size 2) and dense convolution blocks consecutively to extract the compressed encoded feature representation. The dense-convolution blocks from DenseNet \cite{RN222} are used to strengthen feature propagation, encourage feature reuse and substantially reduce the number of parameters. The decoder is composed of deconvolution layers and convolution blocks. The skip connection between the encoder and the decoder side allows for feature reuse and information flow. The architecture has two decoders, one to predict the relevant gland locations, and a second to predicts the gland contours. Thus, the decoders output a gland probability map and a contour probability map. The network is supervised to jointly optimize the prediction of gland locations and gland contours.

\bigbreak
\noindent \textbf{Table 3} Overview of FCN methods

\begin{longtable}{|p{0.3in}|p{0.3in}|p{1.3in}|p{0.7in}|p{0.8in}|p{0.9in}|p{0.7in}|} \hline 
\textbf{Ref.} & \textbf{Year} & \textbf{Network} & \textbf{Supervision} & \textbf{Dimension} & \textbf{Site} & \textbf{Modality} \\ \hline 
\cite{RN90} & 2015 & U-Net & Supervised & 2D slice & Neuronal structure & Electron microscopic \\ \hline 
\cite{RN91} & 2016 & Cascaded FCN & Supervised & 3D volume & Liver and lesion & CT \\ \hline 
\cite{RN249} & 2016 & 3D U-Net & Supervised & 3D volume & Kidney & Xenopus \\ \hline 
\cite{RN111} & 2017 & Dilated FCN & Supervised & 2D slice & Abdomen & CT \\ \hline 
\cite{RN112} & 2017 & 3D FCN Feature Driven Regression Forest & Supervised & 3D patch & Pancreas & CT \\ \hline 
\cite{RN114} & 2017 & 2D FCN & Supervised & 2.5D slices & Whole body & CT \\ \hline 
\cite{RN116} & 2018 & Foveal Fully Convolutional Nets & Supervised & N.A.* & Whole body & CT \\ \hline 
\cite{RN156} & 2018 & DRINet & Supervised & 2D slice & Brain, \newline{abdomen} & CT \\ \hline 
\cite{RN108} & 2018 & 3D U-Net & Supervised & 3D volume & Prostate & MRI \\ \hline 
\cite{RN117} & 2018 & Dense V-Net & Supervised & 3D volume & Abdomen & CT \\ \hline 
\cite{RN118} & 2018 & NiftyNet & Supervised & 3D volume & Abdomen & CT \\ \hline 
\cite{RN119} & 2018 & PU-Net, CU-Net & Supervised & 2D slice & Pelvis & CT \\ \hline 
\cite{RN22} & 2018 & Dilated U-Net & Supervised & 2D slice & Chest & CT \\ \hline 
\cite{RN121} & 2018 & 3D U-JAPA-Net & Supervised & 3D volume & Abdomen & CT \\ \hline 
\cite{RN20} & 2018 & U-Net & Supervised & 2D slice & Pelvis & CT \\ \hline 
\cite{RN122} & 2018 & Cascade 3D FCN & Supervised & 3D patch & Abdomen & CT \\ \hline 
\cite{RN123} & 2018 & Multi-scale Pyramid of 3D FCN & Supervised & 3D patch & Abdomen & CT \\ \hline 
\cite{RN18} & 2018 & Shape representation model constrained FCN & Supervised & 3D volume & Head \& Neck & CT \\ \hline 
\cite{RN128} & 2018 & Hierarchical Dilated Neural Networks & Supervised & 2D slice & Pelvis & CT \\ \hline 
\cite{RN183} & 2019 & Dilated FCN & Supervised & 2D slice & Lung & CT \\ \hline 
\cite{RN130} & 2019 & Dense-U-Net & Supervised & 2D slice & Head \& Neck & Stained colon adenocarcinoma dataset \\ \hline 
\cite{RN163} & 2019 & 2D and 3D FCNs & Supervised & 2D slice and 3D volume & Pulmonary nodule & CT \\ \hline 
\cite{RN132} & 2019 & Dedicated 3D FCN & Supervised & 3D patch & Thorax, \newline{abdomen} & DECT \\ \hline 
\cite{RN23} & 2019 & 2D FCN (DeepLabV3+) & Transfer learning & 2D slice & Pelvis & MRI \\ \hline 
\cite{RN159} & 2019 & 2D FCN & Supervised & 2D patch & Pulmonary vessels & CT \\ \hline 
\cite{RN138} & 2019 & Dual U-Net & Supervised & 2D slice & Glioma Nuclei & Hematoxylin and eosin (H\&E)-stained histopathological image \\ \hline 
\cite{RN141} & 2019 & Consecutive deep encoder-decoder Network & Supervised & 2D slice & Skin lesion & CT \\ \hline 
\cite{RN158} & 2019 & U-Net & Supervised & 2D slice & Lung & HRCT \\ \hline 
\cite{RN162} & 2019 & 3D U-Net & Supervised & 3D volume & Chest & CT \\ \hline 
\cite{RN15} & 2019 & 3D U-Net with Multi-atlas & Supervised & 3D volume & Brain tumor & Dual-energy CT \\ \hline 
\cite{RN147} & 2019 & Triple-Branch FCN & Supervised & N.A. & Abdomen/torso & CT \\ \hline 
\cite{RN28} & 2019 & 2.5D Deeply supervised V-Net & Supervised & 2.5 patch & Prostate & Ultrasound \\ \hline 
\cite{RN34} & 2019 & Group dilated deeply supervised FCN & Supervised & 3D volume & Prostate & MRI \\ \hline 
\cite{RN67} & 2019 & 3D FCN & Supervised & 3D volume & Arteriovenous malformations & Contract-enhanced CT \\ \hline 
\cite{RN51} & 2019 & 3D FCN & Supervised & 3D volume & Left ventricle & SPECT \\ \hline 
\cite{RN63} & 2019 & DeepMAD & Supervised & 2.5D patch & Vessel wall & MRI \\ \hline 
\cite{RN146} & 2019 & 3D U-Net & Supervised & 3D volume & Head \& Neck & CT \\ \hline 
\cite{RN135} & 2019 & OBELISK-Net & Supervised & 3D volume & Abdomen & CT \\ \hline
\end{longtable}

*N.A.: not available, i.e. not explicitly indicated in the publication

\noindent 
\bigbreak
{\bf 3.3.3 Discussion}

The problem of FCN is that the receptive size is fixed so if the object size changes then FCN struggles to detect them all. One solution is multi-scale networks, where input images are resized and fed to the network. Multi-scale techniques can overcome the problem of the fixed receptive size in the FCN \cite{RN123}. However, sharing the parameters of the same network on a resized image is not a very effective way as the object of different scales requires different parameters to process. As another solution for a fixed-size receptive field, for the images with the size bigger than the field of view, the FCN can be applied in a sliding window manner across the entire image \cite{RN167}.
As compared to single FCN architecture, the advantage of cascade FCN is that separate sets of filters can be applied for each stage and therefore the quality of segmentation can be significantly increased. For example, Trullo \textit{et al.} proposed two collaborate FCNs to jointly segment multi-organ in thoracic CT image, one is used for organ localization and the other one is used to segment the organ within that ROI \cite{RN113}. However, because of the additional network involved and two or more steps needed, the computation time of this kind of method would be prolonged as compared to single FCN architecture. In addition, if the first network of cascade FCN is used for organ localization, the performance of this method will largely rely on the accuracy of localization.
Although FCN-based prostate segmentation \cite{RN28, RN34} can reach at good performance, the limitations of this kind of method still exist. First, due to three or more stages of deep supervision and corresponding up-sampling convolutional kernels involved, the computation complexity is higher than the U-Net and V-Net methods. In addition, when using 2.5D patch as input, the segmented contours of three directions may not be well-matched, introducing an adaptive and non-linear contour refinement model, such as conditional random field, would be a future work if researches use this kind of method for multi-organ segmentation. Second, FCNs used voxel-wise loss such as cross entropy for segmentation. However, in the final segmentation map, there is no guarantee of spatial consistency. Recently, in FCN based methods, conditional random filed and graph cut methods are used as segmentation refinement into the FCN-based workflow by incorporating spatial correlation. The limitation of these kinds of segmentation refinement is that they only consider pair-wise potential which might cause boundary leakage in low contrast regions.
\bigbreak

\noindent 
\bigbreak
\subsection{GAN methods}

\noindent 
\bigbreak
{\bf 3.4.1 Network designs}

GANs have gained a lot of attention in medical imaging due to their capability of data generation without explicitly modelling the probability density function. The adversarial loss brought by the discriminator provides a clever way of incorporating unlabeled samples into training and imposing higher order consistency. This has been proven to be useful in many cases, such as image reconstruction \cite{RN232}, image enhancement \cite{RN2, RN21}, segmentation \cite{RN190, RN31}, classification and detection \cite{RN231}, augmentation \cite{RN230}, and cross-modality synthesis \cite{RN23}. 
A typical GAN consists of two competing networks, a generator and a discriminator \cite{RN233}. The generator is trained to generate artificial data that approximate a target data distribution from a low-dimensional latent space. The discriminator is trained to distinguish the artificial data from actual data. The discriminator encourages the generator to predict realistic data by penalizing unrealistic predictions via learning. Therefore, the discriminative loss could be considered as a dynamic network-based loss term. The two networks compete in a zero-sum game. Multiple variants of GAN can be summarized into three categories: 1) variants of discriminator’s objective, 2) variants of generator’s objective, and 3) variants of architecture, which are summarized in \cite{RN235}.

\noindent 
\bigbreak
{\bf 3.4.2 Overview of works}

As discussed above in FCN methods, one challenge in medical image segmentation of traditional FCN methods is that these methods may introduce boundary leakage in low contrast regions. Using adversarial loss introduced via a discriminator can take into account high order potentials to solve this problem \cite{RN189}. The adversarial loss can be regarded as a learned similarity measurement between the segmented contours and the annotated ground truth (manual contours) for medical image segmentation tasks. Instead of only measuring the similarity (such as Dice loss and cross entropy loss) in the voxel domain, the additional discriminator maps the segmented and ground truth contours to a low dimensional feature space to represent the shape information and then uses logistic loss to measure the similarity of the feature vector between segmented contours and manual contours. The idea is similar to the perceptual loss. The difference is that the perceptual loss is computed from a pre-trained classification network on natural images whereas the adversarial loss is computed from a network that trained adaptively during the evolvement of the generator.
Dai \textit{et al.} proposed structure correcting adversarial network (SCAN) to segment lung and the heart in CXR images \cite{RN190}. Compared to FCN, SCAN used an FCN (generator) to generate the binary mask of segmented organs and incorporated a critic network (discriminator) to discriminate the structural regularities emerging from human physiology. During training, the critic network learns to discriminate between the ground truth organ annotations from the masks synthesized by the segmentation network. Through this adversarial process, the critic network is able to learn the higher order structures and to guide the segmentation model to achieve realistic segmentation outcomes.
In medical image multi-organ segmentation, a major limitation of traditional DL-based segmentation methods is their requirement for large amount of paired training image with ground truth contours as learning target. In order to solve this challenge, Mondal \textit{et al.} proposed GAN-based method for 3D multimodal brain MRI segmentation from a few-shot learning perspective \cite{RN197}. The main idea of this work is to leverage the recent success of GANs to train a DL-based model with highly-limited training set of labeled images, without sacrificing the performance of full supervision. The proposed adversarial network encourages the segmentation to have a similar distribution of outputs for images with and without annotations, thereby helping generalization. In addition, few-shot learning method seeks good generalization on problems with very limited labeled dataset, typically containing just a few training samples of the target classes.
Xue \textit{et al.} proposed an adversarial network to train deep neural networks for the segmentation of multiple organs on thoracic CT images \cite{RN31}. The proposed design of adversarial networks, called U-Net-generative-adversarial-network (U-Net-GAN), jointly trains a set of U-Nets as generators and fully convolutional networks (FCNs) as discriminators. U-Net-GAN is a conditional GAN. Specifically, the generator, composed of U-Net, produces image segmentation map of multiple organs by an end-to-end mapping learned from CT image and its labelled organs. The discriminator, structured as FCN, discriminates between the ground truth and segmented organs produced by the generator. The generator and discriminator compete against each other in an adversarial learning process to produce the optimal segmentation map of multiple organs.

\bigbreak
\noindent \textbf{Table 4} Overview of GAN methods

\begin{longtable}{|p{0.3in}|p{0.3in}|p{1.0in}|p{0.8in}|p{0.9in}|p{0.8in}|p{0.9in}|} \hline 
\textbf{Ref.} & \textbf{Year} & \textbf{Network} & \textbf{Supervision} & \textbf{Dimension} & \textbf{Site} & \textbf{Modality} \\ \hline 
\cite{RN190} & 2015 & SCAN & Supervised & 2D slice & Chest & X-rays \\ \hline 
\cite{RN192} & 2017 & Multi-connected adversarial networks & Unsupervised & 2D slice & Brain & Multi-modality MRI \\ \hline 
\cite{RN187} & 2017 & Dilated GAN & Supervised & 2D slice & Brain & MRI \\ \hline 
\cite{RN194} & 2017 & Conditional GAN & Supervised & 2D slice & Brain tumor & MRI \\ \hline 
\cite{RN188} & 2017 & GAN & Supervised & 2D patch & Retinal Vessel & Fundoscopic  \\ \hline 
\cite{RN189} & 2017 & Adversarial Image-to-Image Network & Supervised & 3D volume & Liver & CT \\ \hline 
\cite{RN191} & 2017 & Adversarial FCN-CRF Nets & Supervised & 2D slice & Mass & Mammograms \\ \hline 
\cite{RN195} & 2018 & GAN & Supervised & N.A.* & Brain tumor & MRI \\ \hline 
\cite{RN197} & 2018 & Few-shot GAN & Semi-supervised & 3D patch & Brain & MRI \\ \hline 
\cite{RN193} & 2018 & Context-aware GAN & Supervised & 2D cropped slices & Cardiac & MRI \\ \hline 
\cite{RN196} & 2018 & Conditional Generative Refinement Adversarial Networks & Supervised & 2D slice & Brain & MRI \\ \hline 
\cite{RN186} & 2018 & SegAN & Supervised & 2D slice & Brain & MRI \\ \hline 
\cite{RN198} & 2018 & MDAL & Supervised & 2D slice & Left and Right-Ventricular & Cardiac MRI \\ \hline 
\cite{RN127} & 2018 & TD-GAN & Unsupervised & 2D slice & Whole body & X-ray \\ \hline 
\cite{RN31} & 2019 & U-Net-GAN & Supervised & 3D volume & Thorax & CT \\ \hline 
\cite{RN139} & 2019 & Conditional GAN & Supervised & 2D slice & Nuclei & Histopathology Images \\ \hline 
\cite{RN145} & 2019 & Distance-aware GAN & Supervised & 2D slice & Chest & CT \\ \hline 
\end{longtable}

*N.A.: not available, i.e. not explicitly indicated in the publication

\noindent 
\bigbreak
{\bf 3.4.3 Discussion}

In segmentation tasks, GAN is efficient in prediction stage, since it only needs to perform a forward pass through the generator network (for segmentation). Using adversarial loss as a shape regulator can benefit more when the learning target (organ) has a regular shape, e.g., for lung and heart, but would be less useful for other small tubular objects, such as vessels and catheters.
In \cite{RN31}, GAN was applied to delineate the left and right lungs, spinal cord, esophagus, and heart using 35 patients’ chest CTs. The averaged DSC for the above five OARs are 0.97, 0.97, 0.90, 0.75 and 0.87, respectively. The mean surface distance of the five OARs obtained with GAN method ranges between 0.4 mm and 1.5 mm on average among all 35 patients. The mean dose differences on the 20 SBRT lung plans using the segmented results ranged from -0.001 to 0.155 Gy for the five OARs. This demonstrates that GAN is a potentially valuable method for improving the efficiency of the lung radiotherapy treatment planning.
GAN methods have realistic indications. Patient movement, such as translation and rotation, doesn’t change the relative position among organs. Including the transformed data could help avoid overfitting and help the segmentation algorithm learn this invariant property. However, for multi-organ segmentation, due to the size and shape differences among different organs and variation among patients, it is difficult to balance the loss function among the different organs. Integrating all the segmentations into one network complicates the training process and reduces segmentation accuracy. To simplify the method, previous GAN method \cite{RN31} grouped OARs of similar dimensions, and utilized three sub-networks for segmentation, one for lungs and heart, and the other two for esophagus and spinal cord, respectively. This approach improves segmentation accuracy at the cost of computation efficiency. But it also introduces additional computation time for both training and prediction. It could be an issue if the task is to apply the GAN method to segment more OARs simultaneously. Simultaneously determining the location of organs and segmenting the organs within that location for multi-organ segmentation would be a future work, i.e., the need is to explore the possibility of multi-organ segmentation in one compound network.

\noindent 
\bigbreak
\subsection{R-CNN methods}
\noindent
\bigbreak
{\bf 3.5.1 Network designs}

In medical image multi-organ segmentation, as we discussed above, simultaneous segmenting multi-organ is challenging, because it requires the correct detection of all organs in an image volume while also accurately segment the organs within that detection. It is similar to the classical computer vision tasks of instance segmentation, which include two subtasks: one is the object detection with the goal of classifying individual objects and localizing each using a bounding box (ROI to medical image), the other one is the semantic segmentation with the goal of classifying each pixel into a fixed set of categories without differentiating object instances. Recently, the development of region-CNN (R-CNN) family introduced a simple and flexible way to solve this challenge.
R-CNN is a network based on work with regions \cite{RN236}. To bypass the problem of selecting a large number of regions, R-CNN utilized a selective search \cite{RN237} to extract 2000 candidate regions from image. These regions were called as region proposals. By warping to a same size, these regional proposals were then fed into a CNN to extract a 4096-dimensional feature vector as output. The CNN acts as a feature extractor. The output dense layer consists of the features extracted from the image and the extracted features are fed into an SVM to classify the presence of the object within that region proposal. In addition to predicting the presence of an object within the region proposals, the algorithm also predicts four values (2D version) which are offset values to increase the precision of the bounding box.
R-CNN needs large computation time to train the network with 2000 region proposals per 2D image slice. To address this issue, Girshick \textit{et al.} proposed a faster objection algorithm called Fast R-CNN \cite{RN238}. Compared to R-CNN, instead of selecting region proposals and feeding them into a CNN, the region proposals were obtained via first feeding original image into an FCN (called as backbone) to obtain the convolutional feature map to identify the regional proposals, then warping them into squares and reshaping them into a fixed size via a ROI pooling layer. By using a fully connected layer, the regional proposal was projected to an ROI feature vector. Finally, softmax layer was used to predict the class of that region proposal and also the offset values for the bounding box. The reason that Fast R-CNN is faster than R-CNN is because it does not need to feed 2000 region proposals to the CNN for each feeding image. Instead, the convolution operation is done only once per image and a feature map is generated from it.
 Both R-CNN and Fast R-CNN use selective search to identify the region proposals. However, selective search is time-consuming. To solve this problem, Ren \textit{et al.} proposed an object detection algorithm that eliminates the selective search and lets the network learn the regional proposals, called as Faster R-CNN \cite{RN239}. Similar to Fast R-CNN, Faster R-CNN first fed the image into an FCN to extract convolutional feature map. Instead of using selective search on the feature map to identify the region proposals, a separate network was used to predict the region proposals. The predicted region proposals were then reshaped using a ROI pooling layer which was then used to classify the image within the proposed region and predict the offset values for the bounding boxes.
Based on the ground works of feature extraction and regional proposals identification built by Faster R-CNN, performing image segmentation within the detected bounding box (ROI) is easy to achieve. After ROI pooling layer in Faster R-CNN, He \textit{et al.} integrated 2 more convolution layers to build the semantic segmentation within the ROI, called as Mask R-CNN \cite{RN240}. Another major contribution of Mask R-CNN is the refinement of the ROI pooling. In previous Faster R-CNN, Fast R-CNN and R-CNN methods the ROI warping is digitalized: the cell boundaries of the target feature map are forced to realign with the boundary of the input feature maps. Therefore, each target cells may not be in the same size. Mask R-CNN uses ROI Align which does not digitalize the boundary of the cells and make every target cell to have the same size. It also applies interpolation to better calculate the feature map values within the cell.

\noindent 
\bigbreak
{\bf 3.5.2 Overview of works}

In order to solve the problem of low-quality of CT image, the lack of annotated data, and the complex shapes of lung nodules, Liu \textit{et al.} applied 2D Mask R-CNN for lung pulmonary nodule segmentation in a transfer learning manner \cite{RN203}. The Mask R-CNN is trained on the COCO data set, which is a natural image dataset, and was then fine-tuned to segment pulmonary nodules. As an improvement, Kopelowitz and Engelhard applied 3D Mask R-CNN to handle 3D CT image volume to detect and segment the lung nodules \cite{RN201}. 
Xu \textit{et al.} proposed a novel heart segmentation pipeline which combined Faster R-CNN and U-Net, abbreviated by CFUN \cite{RN199}. Due to Faster R-CNN’s precise localization ability and U-Net’s powerful segmentation ability, CFUN needs only one-step detection and segmentation inference to get the whole heart segmentation result, obtaining good results with significantly reduced computational cost. Besides, CFUN adopts a new loss function based on edge information named 3D Edge-loss as an auxiliary loss to accelerate the convergence of training and improve the segmentation results. Extensive experiments on the public dataset show that CFUN exhibits competitive segmentation performance in a sharply reduced inference time. Similarly, Bouget \textit{et al.} proposed a combination of Mask R-CNN and U-Net for the segmentation and detection of mediastinal lymph nodes and anatomical structures in CT data for lung cancer staging \cite{RN164}.
Early detection of lung cancer is crucial in reducing mortality. MRI may be a viable imaging technique for lung cancer detection. Li \textit{et al.} proposed a lung nodule detection method based on Faster R-CNN for thoracic MRI in a transfer learning manner \cite{RN206}. A false positive (FP) reduction scheme based on anatomical characteristics is designed to reduce FPs and preserve the true nodule.  Similarly, Faster R-CNN was also used for pulmonary nodule detection on CT image \cite{RN176}.
Xu \textit{et al.} proposed an efficient detection method for multi-organ localization in CT image using 3D regional proposal network (RPN) \cite{RN200}. Since the proposed RPN is implemented in a 3D manner, it can take advantage of the spatial context information in CT image. AlexNet was used to build the backbone network architecture that is able to generate high-resolution feature maps to further improve the localization performance of small organs. The method was evaluated on abdomen and brain site datasets, and achieved high detection precision and localization accuracy with fast inference speed.

\bigbreak
\noindent \textbf{Table 5} Overview of RPN methods

\begin{longtable}{|p{0.3in}|p{0.3in}|p{1.2in}|p{0.8in}|p{0.7in}|p{1.2in}|p{0.6in}|} \hline 
\textbf{Ref.} & \textbf{Year} & \textbf{Network} & \textbf{Supervision} & \textbf{Dimension} & \textbf{Site} & \textbf{Modality} \\ \hline 
\cite{RN203} & 2018 & Mask R-CNN & Transfer learning & 2D slice & Lung nodule & CT \\ \hline 
\cite{RN199} & 2018 & Combination of Faster R-CNN and U-Net (CFUN) & Supervised & 3D volume & Cardiac & CT \\ \hline 
\cite{RN164} & 2019 & Combination of U-Net and Mask R-CNN & Supervised & 2D slice & Chest & CT \\ \hline 
\cite{RN176} & 2019 & Faster R-CNN & Supervised & 2D slice & Thorax/pulmonary nodule & CT \\ \hline 
\cite{RN201} & 2019 & 3D Mask R-CNN & Supervised & 3D volume & Lung nodule & CT \\ \hline 
\cite{RN206} & 2019 & 3D Faster R-CNN & Supervised & 3D volume & Thorax/ lung nodule & MRI \\ \hline 
\cite{RN202} & 2019 & Mask R-CNN & Supervised & N.A.* & Chest & X-Ray \\ \hline 
\cite{RN200} & 2019 & 3D RPN & Supervised & 3D volume & Whole body & CT \\ \hline 
\cite{RN204} & 2019 & Multiscale Mask R-CNN & Supervised & 2D slice & Lung tumor & PET \\ \hline 
\end{longtable}

*N.A.: not available, i.e. not explicitly indicated in the publication

\noindent 
\bigbreak
{\bf 3.5.3 Discussion}

In the work of \cite{RN203}, researchers used Mask R-CNN to segment lung nodules for the first time. After a series of comparative experiments, ResNet101 and feature pyramid network (FPN) were selected as the backbone of Mask R-CNN. Experimental results showed that it not only located the location of nodules, but also provided nodule contour information. It provided more detailed information for cancer treatment. The proposed method was validated on the LIDC-IDRI data set and achieved desired accuracy. However, due to the 2D network design, the spatial information of CT image will be lost. 3D contexts play an import role in recognizing nodules. A 3D Mask R-CNN would perform better than a 2D as it also captures the vertical information.
For the detection of lung nodule using Faster R-CNN, limitations still exist. First, small and low contrast nodules are not successfully detected by Faster R-CNN. This challenging may also occur for other multi-organ segmentation when small organs exist, such as esophagus in lung segmentation. Second, researchers find that some air artifacts and juxta heart tissues may be falsely detected as nodules. In order to alleviate these problems, Li \textit{et al.} design a filter to improve the image quality and remove these artifacts \cite{RN206}. In addition, multi-scale strategy was introduced in the whole detection system to increase the detection rate of small and low contrast nodules.
R-CNN family methods could be efficient tools for several multi-organ segmentation and detection tasks. However, technical adjustments and optimizations may be required to make the extended model to achieve comparable performance to the methods dedicated to organ segmentation. Due to the higher data dimensionality and larger number of weight parameters, training 3D R-CNN based models is more time-consuming than the 2D version. However, significant advantages, such as higher localization accuracy and much higher prediction speed, still encourage us to handle this problem using 3D R-CNNs. To speed up the training procedure of the proposed method, one potential solution is to apply batch normalization after each convolutional layer in the backbone network to improve the model convergence, and conduct most calculations on GPU in parallel \cite{RN200}. 

\noindent 
\bigbreak
\subsection{Hybrid methods}
\noindent 
\bigbreak
{\bf 3.6.1 Network designs}

Due to the poor image quality, such as low contrast around organ boundary, recently some methods used hybrid designs to solve this challenge. The hybrid design usually involves two or more networks for different functional propose. For example, one network aims to enhance the image quality, and the other one to segment the OARs from the enhanced image. We think that this will be a new trend for multi-organ segmentation.

\noindent 
\bigbreak
{\bf 3.6.2 Overview of works}

Accurate segmentation of the pelvic OARs on CT image for treatment planning is challenging due to the poor soft-tissue contrast \cite{RN46, RN54}. MRI has been used to aid prostate delineation, but its final accuracy is limited by MRI-CT registration errors \cite{RN32, RN52}. Lei \textit{et al.} developed a deep attention-based segmentation strategy on CT-based synthetic MRI (sMRI) via cycle generative adversarial network (CycleGAN) \cite{RN23} to deal with the low contrast soft tissue organ (such as bladder, prostate and rectum) delineation challenge without MRI acquisition \cite{RN7}. This hybrid method includes two main steps: first, a CycleGAN was used to estimate sMRI from CT images. Second, a deep attention FCN was trained based on sMRI and manual contours deformed from MRIs. Attention models were introduced to pay more attention to prostate boundary. Inspired by this method, Xue \textit{et al.} developed a novel sMRI-aided segmentation method for mail pelvic CT multi-organ segmentation \cite{RN4}. Similarly, Lei \textit{et al.} introduced this kind of method for the multi-organ segmentation of cone-beam computed tomography (CBCT) pelvic data for potential CBCT-guided adaptive radiation therapy workflow \cite{RN3}.
In multi-organ segmentation of abdominal CT scans, as we discussed above, supervised DL-based algorithms require lots of voxel-wise annotations, which are usually difficult, expensive, and slow to obtain. However, massive unlabeled 3D CT volumes are usually easily accessible. Zhou \textit{et al.} proposed Deep Multi-Planar CoTraining (DMPCT) in a semi-supervised learning manner to solve this problem \cite{RN150}. The DMPCT network architecture includes three steps: 1) A DL-based network is learned in a co-training manner to mine consensus information from 2D patches extracted from multiple planes. The DL-based network is called as teacher model in their work. 2) The trained teacher model is then used to assign pseudo labels to the unlabeled data. Multiplanar fusion is applied to generate more reliable pseudo labels to alleviates the errors occurring in the pseudo labels and thus can help train better segmentation networks. 3) An additional network, called student model, is trained on the union of the manual labeled data and automatically labeled data (called as self-labeled samples) to enlarge the data variation of the training data.

\bigbreak
\noindent \textbf{Table 6} Overview of hybrid methods

\begin{longtable}{|p{0.5in}|p{0.3in}|p{1.4in}|p{0.9in}|p{0.7in}|p{0.6in}|p{0.6in}|} \hline 
\textbf{Ref.} & \textbf{Year} & \textbf{Network} & \textbf{Supervision} & \textbf{Dimension} & \textbf{Site} & \textbf{Modality} \\ \hline 
\cite{RN4, RN7} & 2019 & Synthetic MRI-aided FCN & Supervised & 2.5D patch & Pelvic & CT \\ \hline 
\cite{RN3} & 2019 & Synthetic MRI-aided deep attention FCN & Supervised & 3D volume & Pelvic & CBCT \\ \hline 
\cite{RN150} & 2019 & Deep Multi-Planar Co-Training (DMPCT) & Co-training & 3D volume & Abdomen & CT \\ \hline 
\end{longtable}

\noindent 
\bigbreak
{\bf 3.6.3 Discussion}

Compared to CT and CBCT images, the superior soft-tissue contrast of sMRI improves the prostate segmentation accuracy and alleviate the issue of prostate volume overestimation when using CT images alone. However, in sMRI-aided segmentation methods, the registration between training MRI and CT or CBCT will affect the sMRI image quality and thus finally affect the segmentation network performance. In this sense, the registration error also affects the delineation accuracy ultimately. Thus, this kind of method relies on the accurate deformable registration.
Since hybrid methods do not need a large number of training multi-modality data collection to provide comprehensive information or a large number of manual delineated contours collection for learning target (the annotation of multiple organs in 3D volumes requires massive labor from radiologists), hybrid methods can be practical for clinical applications. To demonstrate hybrid methods’ performance in multiple complex anatomical structures, such as thorax segmentation, will be a future work.

\noindent 
\bigbreak
\section{ Benchmark}
Benchmarking is important for readers to understand through comparing the performance of multiple methods.

\noindent 
\bigbreak
\subsection{2017 AAPM Thoracic Auto-segmentation Challenge}
One benchmark dataset used in this work is from 2017 AAPM Thoracic Auto-segmentation Challenge \cite{RN241}, which provide a benchmark dataset and platform for evaluating performance of automatic multi-organ segmentation methods of in thoracic CT images. The OARs include left and right lungs, heart, esophagus, and spinal cord. Sixty thoracic CT scans provided by three different institutions were separated into 36 training, 12 offline testing, and 12 online testing scans. Clinical contours used for treatment planning were quality checked and edited to adhere to the RTOG 1106 contouring guidelines. The Dice similarity coefficient (DSC), 95\% Hausdorff distance (HD) and mean surface distance (MSD) metrics were used to evaluate the competing algorithms’ performance. 

From the report of this challenge, there are 7 participants completed the online challenge. 5 out of 7 participants used DL-based methods. In addition to reporting the DL-based methods participated in this challenge, we also searched and combined the numerical report of chest multi-organ segmentation in recent studies using this benchmark data. The numerical results of different DL-based methods are listed as follows:

\bigbreak
\noindent \textbf{Table 7} The comparison of the results from DL-based methods using datasets from 2017 AAPM Thoracic Auto-segmentation Challenge.

\begin{longtable}{|p{0.5in}|p{1.0in}|p{0.7in}|p{0.7in}|p{0.7in}|p{0.7in}|p{0.7in}|} \hline 
\textbf{Metric} & \textbf{Method} & \textbf{Esophagus} & \textbf{Heart} & \textbf{Left Lung} & \textbf{Right Lung} & \textbf{Spinal Cord} \\ \hline 
\multirow{6}{*}\newline{\textbf{DSC}}
& DCNN\newline Team Elekta & 0.72$\mathrm{\pm}$0.10 & 0.93$\mathrm{\pm}$0.02 & 0.97$\mathrm{\pm}$0.02 & 0.97$\mathrm{\pm}$0.02 & 0.88$\mathrm{\pm}$0.04 \\ \cline{2-7}
& 3D U-Net \cite{RN242} & 0.72$\mathrm{\pm}$0.10 & 0.93$\mathrm{\pm}$0.02 & 0.97$\mathrm{\pm}$0.02 & 0.97$\mathrm{\pm}$0.02 & 0.89$\mathrm{\pm}$0.04 \\ \cline{2-7}
& Multi-class CNN\newline Team Mirada & 0.71$\mathrm{\pm}$0.12 & 0.91$\mathrm{\pm}$0.02 & 0.98$\mathrm{\pm}$0.02 & 0.97$\mathrm{\pm}$0.02 & 0.87$\mathrm{\pm}$0.11 \\ \cline{2-7}
& 2D ResNet\newline Team Beaumont & 0.61$\mathrm{\pm}$0.11 & 0.92$\mathrm{\pm}$0.02 & 0.96$\mathrm{\pm}$0.03 & 0.95$\mathrm{\pm}$0.05 & 0.85$\mathrm{\pm}$0.04 \\ \cline{2-7}
& 3D and 2D U-Net\newline Team WUSTL & 0.55$\mathrm{\pm}$0.20 & 0.85$\mathrm{\pm}$0.04 & 0.95$\mathrm{\pm}$0.03 & 0.96$\mathrm{\pm}$0.02 & 0.83$\mathrm{\pm}$0.08 \\ \cline{2-7}
& U-Net-GAN \cite{RN31} & 0.75$\mathrm{\pm}$0.08 & 0.87$\mathrm{\pm}$0.05 & 0.97$\mathrm{\pm}$0.01 & 0.97$\mathrm{\pm}$0.01 & 0.90$\mathrm{\pm}$0.04 \\ \cline{1-7}
\multirow{6}{*}\newline{\textbf{MSD\newline (mm)}}
& DCNN\newline Team Elekta & 2.23$\mathrm{\pm}$2.82 & 2.05$\mathrm{\pm}$0.62 & 0.74$\mathrm{\pm}$0.31 & 1.08$\mathrm{\pm}$0.54 & 0.73$\mathrm{\pm}$0.21 \\ \cline{2-7} 
& 3D U-Net \cite{RN242} & 2.34$\mathrm{\pm}$2.38 & 2.30$\mathrm{\pm}$0.49 & 0.59$\mathrm{\pm}$0.29 & 0.93$\mathrm{\pm}$0.57 & 0.66$\mathrm{\pm}$0.25 \\ \cline{2-7} 
& Multi-class CNN\newline Team Mirada & 2.08$\mathrm{\pm}$1.94 & 2.98$\mathrm{\pm}$0.93 & 0.62$\mathrm{\pm}$0.35 & 0.91$\mathrm{\pm}$0.52 & 0.76$\mathrm{\pm}$0.60 \\ \cline{2-7}
& 2D ResNet\newline Team Beaumont & 2.48$\mathrm{\pm}$1.15 & 2.61$\mathrm{\pm}$0.69 & 2.90$\mathrm{\pm}$6.94 & 2.70$\mathrm{\pm}$4.84 & 1.03$\mathrm{\pm}$0.84 \\ \cline{2-7}
& 3D and 2D U-Net\newline Team WUSTL & 13.10$\mathrm{\pm}$10.39 & 4.55$\mathrm{\pm}$1.59 & 1.22$\mathrm{\pm}$0.61 & 1.13$\mathrm{\pm}$0.49 & 2.10$\mathrm{\pm}$2.49 \\ \cline{2-7}
& U-Net-GAN \cite{RN31} & 1.05$\mathrm{\pm}$0.66 & 1.49$\mathrm{\pm}$0.85 & 0.61$\mathrm{\pm}$0.73 & 0.65$\mathrm{\pm}$0.53 & 0.38$\mathrm{\pm}$0.27 \\ \cline{1-7}
\multirow{6}{*}\newline{\textbf{HD95\newline (mm)}}
& DCNN\newline Team Elekta & 7.3+10.31 & 5.8$\mathrm{\pm}$1.98 & 2.9$\mathrm{\pm}$1.32 & 4.7$\mathrm{\pm}$2.50 & 2.0$\mathrm{\pm}$0.37 \\ \cline{2-7}
& 3D U-Net \cite{RN242} & 8.71+10.59 & 6.57$\mathrm{\pm}$1.50 & 2.10$\mathrm{\pm}$0.94 & 3.96$\mathrm{\pm}$2.85 & 1.89$\mathrm{\pm}$0.63 \\ \cline{2-7}
& Multi-class CNN\newline Team Mirada & 7.8$\mathrm{\pm}$8.17 & 9.0$\mathrm{\pm}$4.29 & 2.3$\mathrm{\pm}$1.30 & 3.7$\mathrm{\pm}$2.08 & 2.0$\mathrm{\pm}$1.15 \\ \cline{2-7}
& 2D ResNet\newline Team Beaumont & 8.0$\mathrm{\pm}$3.80 & 8.8$\mathrm{\pm}$5.31 & 7.8$\mathrm{\pm}$19.13 & 14.5$\mathrm{\pm}$34.4 & 2.3$\mathrm{\pm}$0.50 \\ \cline{2-7}
& 3D and 2D U-Net\newline Team WUSTL & 37.0$\mathrm{\pm}$26.88 & 13.8$\mathrm{\pm}$5.49 & 4.4$\mathrm{\pm}$3.41 & 4.1$\mathrm{\pm}$2.11 & 8.10$\mathrm{\pm}$10.72 \\ \cline{2-7}
& U-Net-GAN \cite{RN31} & 4.52$\mathrm{\pm}$3.81 & 4.58$\mathrm{\pm}$3.67 & 2.07$\mathrm{\pm}$1.93 & 2.50$\mathrm{\pm}$3.34 & 1.19$\mathrm{\pm}$0.46 \\ \cline{1-7}
\end{longtable}

*Note: Methods without followed reference were directly from AAPM thorax challenge report \cite{RN241}.

\noindent 
\bigbreak
\subsection{2015 MICCAI Head and Neck Auto-segmentation Challenge}
The other benchmark dataset used in this work is from 2015 MICCAI Head and Neck Auto-segmentation Challenge \cite{RN244}, which provide a benchmark dataset and platform for evaluating performance of automatic multi-organ segmentation methods of in head \& neck CT images. The OARs include nine structures in the head and neck region of CT images: brainstem, mandible, chiasm, bilateral optic nerves, bilateral parotid glands, and bilateral submandibular glands. For this challenge data, a subset of 40 images was used: 25 images were used as training data, 10 images were used for off-site testing, and 5 images were used for on-site testing. The subset was chosen to ensure that all structures were completely included within the CT images, image quality was adequate, and that structures minimally overlapped tumor volumes. No restriction with respect to age or gender was made. 
However, from the report of this challenge \cite{RN244}, there is no team applying DL-based method for that challenge data. We also searched recent DL-based multi-organ segmentation studies that used this data to test their algorithms performance and list state-of-the-art results in the following Table 8. DSC and HD95 were used to evaluate the performance of competing algorithms.

\bigbreak
\noindent \textbf{Table 8} The comparison of the results from DL-based methods using datasets from 2015 MICCAI Head and Neck Auto-segmentation Challenge.

\begin{longtable}{|p{0.4in}|p{0.6in}|p{0.6in}|p{0.6in}|p{0.6in}|p{0.6in}|p{0.6in}|p{0.6in}|p{0.6in}|} \hline 
\textbf{Metric} & \textbf{Organs} & \textbf{Shape model constrained FCN }\cite{RN244} & \textbf{Two-stage U-Net }\cite{RN246} & \textbf{Anatomy Net }\cite{RN247} & \textbf{DL-based }\cite{RN248} & \textbf{Synthetic MRI-aided }\cite{RN71} & \textbf{3D U-Net }\cite{RN249} & \textbf{3D-CNN }\cite{RN250} \\ \hline 
\multirow{9}{*}\newline{\textbf{DSC}}
& \textbf{Brain Stem} &0.87$\mathrm{\pm}$0.03 & 0.88$\mathrm{\pm}$0.02 & 0.87$\mathrm{\pm}$0.02 & 0.87$\mathrm{\pm}$0.03 & 0.91$\mathrm{\pm}$0.02 & 0.80$\mathrm{\pm}$0.08 & N.A. \\ \cline{2-9}
& \textbf{Chiasm} & 0.58$\mathrm{\pm}$0.1 & 0.45$\mathrm{\pm}$0.17 & 0.53$\mathrm{\pm}$0.15 & 0.62$\mathrm{\pm}$0.1 & 0.73$\mathrm{\pm}$0.11 & N.A. & 0.58$\mathrm{\pm}$0.17 \\  \cline{2-9} 
& \textbf{Mandible} & 0.87$\mathrm{\pm}$0.03 & 0.93$\mathrm{\pm}$0.02 & 0.93$\mathrm{\pm}$0.02 & 0.95$\mathrm{\pm}$0.01 & 0.96$\mathrm{\pm}$0.01 & 0.94$\mathrm{\pm}$0.02 & N.A. \\  \cline{2-9}
& \textbf{Left Optic Nerve} & 0.65$\mathrm{\pm}$0.05 & 0.74$\mathrm{\pm}$0.15 & 0.72$\mathrm{\pm}$0.06 & 0.75$\mathrm{\pm}$0.07 & 0.78$\mathrm{\pm}$0.09 & 0.72$\mathrm{\pm}$0.06 & 0.72$\mathrm{\pm}$0.08 \\ \cline{2-9}
& \textbf{Right Optic Nerve} & 0.69$\mathrm{\pm}$0.5 & 0.74$\mathrm{\pm}$0.09 & 0.71$\mathrm{\pm}$0.1 & 0.72$\mathrm{\pm}$0.06 & 0.78$\mathrm{\pm}$0.11 & 0.70$\mathrm{\pm}$0.07 & 0.70$\mathrm{\pm}$0.09 \\ \cline{2-9} 
& \textbf{Left Parotid} & 0.84$\mathrm{\pm}$0.02 & 0.86$\mathrm{\pm}$0.02 & 0.88$\mathrm{\pm}$0.02 & 0.89$\mathrm{\pm}$0.02 & 0.88$\mathrm{\pm}$0.04 & 0.87$\mathrm{\pm}$0.03 & N.A. \\ \cline{2-9} 
& \textbf{Right Parotid} & 0.83$\mathrm{\pm}$0.02 & 0.85$\mathrm{\pm}$0.07 & 0.87$\mathrm{\pm}$0.04 & 0.88$\mathrm{\pm}$0.05 & 0.88$\mathrm{\pm}$0.06 & 0.85$\mathrm{\pm}$0.07 & N.A. \\ \cline{2-9}
& \textbf{Left Submandibular} & 0.76$\mathrm{\pm}$0.06 & 0.76$\mathrm{\pm}$0.15 & 0.81$\mathrm{\pm}$0.04 & 0.82$\mathrm{\pm}$0.05 & 0.86$\mathrm{\pm}$0.08 & 0.76$\mathrm{\pm}$0.09 & N.A. \\ \cline{2-9}
& \textbf{Right Submandibular} & 0.81$\mathrm{\pm}$0.06 & 0.73$\mathrm{\pm}$0.01 & 0.81$\mathrm{\pm}$0.04 & 0.82$\mathrm{\pm}$0.05 & 0.85$\mathrm{\pm}$0.10 & 0.78$\mathrm{\pm}$0.07 & N.A. \\ \cline{1-9}
\multirow{9}{*}\newline{\textbf{HD95\newline{(mm)}}} 
& \textbf{Brain Stem} & 4.01$\mathrm{\pm}$0.93 & 2.01$\mathrm{\pm}$0.33 & N.A. & N.A. & N.A. & N.A. & N.A. \\ \cline{2-9}
& \textbf{Chiasm} & 2.17$\mathrm{\pm}$1.04 & 2.83$\mathrm{\pm}$1.42 & N.A. & N.A. & N.A. & N.A. & 2.81$\mathrm{\pm}$1.56 \\ \cline{2-9}
& \textbf{Mandible} & 1.50$\mathrm{\pm}$0.32 & 1.26$\mathrm{\pm}$0.50 & N.A. & N.A. & N.A. & N.A. & N.A. \\ \cline{2-9}
& \textbf{Left Optic Nerve} & 2.52$\mathrm{\pm}$1.04 & 2.53$\mathrm{\pm}$2.34 & N.A. & N.A. & N.A. & N.A. & 2.33$\mathrm{\pm}$0.84 \\ \cline{2-9}
& \textbf{Right Optic Nerve} & 2.90$\mathrm{\pm}$1.88 & 2.13$\mathrm{\pm}$2.45 & N.A. & N.A. & N.A. & N.A. & 2.13$\mathrm{\pm}$0.96 \\ \cline{2-9}
& \textbf{Left Parotid} & 3.97$\mathrm{\pm}$2.15 & 2.41$\mathrm{\pm}$0.54 & N.A. & N.A. & N.A. & N.A. & N.A. \\ \cline{2-9}
& \textbf{Right Parotid} & 4.20$\mathrm{\pm}$1.27 & 2.93$\mathrm{\pm}$1.48 & N.A. & N.A. & N.A. & N.A. & N.A. \\ \cline{2-9}
& \textbf{Left Submandibular} & 5.59$\mathrm{\pm}$3.93 & 2.86$\mathrm{\pm}$1.60 & N.A. & N.A. & N.A. & N.A. & N.A. \\ \cline{2-9}
& \textbf{Right Submandibular} & 4.84$\mathrm{\pm}$1.67 & 3.44$\mathrm{\pm}$1.55 & N.A. & N.A. & N.A. & N.A. & N.A. \\ \cline{1-9}
\end{longtable}

\noindent *N.A.: not available, i.e. not explicitly indicated in the publication

\noindent 
\bigbreak
\section{ Challenges and Trends}

\noindent 
\bigbreak
\subsection{Over-fitting}
One limitation in multi-organ segmentation is data scarcity, usually leading to overfitting which refers to a model that has good performance on the training dataset but not new data. Large number of labels for training are often not available, because expert manual delineation is time-consuming, laborious and sometimes error prone. When training complex neural networks with limited training data, some specific adjustment should be integrated into the network to prevent the over-fitting. 

\noindent 
\bigbreak
\subsection{Class imbalance}
Another challenge in multi-organ segmentation is class imbalanced. For example, for segmentation of thorax CT organs, the esophagus and spinal cord are often much smaller than the lung. Training a network with class imbalanced data would cause an unstable segmentation model, which is biased towards the classes of large organs. As a result, the choice of loss functions is crucial for these tasks. Another option is generating synthetic samples and integrate these samples into the training data. However, the synthetic samples could reduce learning efficiency since they contain redundant information as the training dataset.

\noindent 
\bigbreak
\subsection{Computational complexity}
The complexity of a DL-based models is defined by input image size, network structure and the number of learnable parameters. In order to avoid the GPU memory limitation and speed up the segmentation, one can reduce the layers or parameters, and focus on methods that artificially increase the number of training data instead of changing the network architecture.

\noindent 
\bigbreak
\subsection{Inter- and intra-observer error}

During training, the ground truth contours were obtained by manual delineation by physicians. In the scenario that manual contouring would be performed differently by different person, or same person at different situation, there is a chance that the DL-based method would be biased toward physicians contouring style as system error, and contouring uncertainty as random error. However, this limitation would be expected to exist in all supervised learning-based methods.

\noindent 
\bigbreak
\subsection{Low image quality}

Images’ inferior quality, such as low contrast around organ boundary, scatter or noise artifact, inhomogeneity would affect the multi-organ segmentation accuracy. As we discussed above, several techniques can be used to reduce these effects, such as deep supervision, deep attention and synthetic image-aided techniques.

\noindent 
\bigbreak
\section{ Conclusion}

This review work covers the current state-of-the-art deep learning-based auto-segmentation approaches in various medical applications. Thoracic contouring challenge at AAPM 2017 and head \& neck contouring challenge at MICCAI 2015 are used as benchmark to show the performance of recent deep learning-based methods.
Judging from the statistics of the cited works, there is a clear trend of using fully convolution network to perform end-to-end semantic segmentation for multi-organ automatic delineating. Recently, GAN-based methods have been used to enhance the reliability of segmented contour. We should see a steady growth of GAN-based multi-organ segmentation. CNN and hybrid methods start to gain popularity in medical image segmentation, although only a few were applicable multi-organ segmentation. We speculate that more research will be focused on R-CNN and hybrid methods in the future.

\noindent 
\bigbreak
{\bf ACKNOWLEDGEMENT}

This research was supported in part by the National Cancer Institute of the National Institutes of Health under Award Number R01CA215718 and Emory Winship Cancer Institute pilot grant.

\noindent 
\bigbreak
{\bf Disclosures}

The authors declare no conflicts of interest.

\noindent 

\bibliographystyle{plainnat}  
\bibliography{manuscript}       

\end{document}